\documentclass[manuscript]{acmart}
\usepackage{textgreek}
\AtBeginDocument{%
  }

\setcopyright{none}
\copyrightyear{2026}
\acmYear{2026}
\acmDOI{}
\acmConference[CHI '27]{CHI Conference on Human Factors in Computing Systems}{April 2027}{}{}
\acmISBN{978-1-4503-XXXX-X/2018/06}

\usepackage{multirow}
\begin{document}

\title{What Is the Role for Humans in Creative Writing? Creativity and Diversity in Human–AI Collaboration}


\author{Chenyi Li}
\correspondingauthor
\affiliation{%
  \institution{Cornell University}
  \city{Ithaca}
  \country{USA}}
\email{cl2836@cornell.edu}

\author{Haoyu Hu}
\affiliation{%
  \institution{Cornell University}
  \city{Ithaca}
  \country{USA}}
\email{hh824@cornell.edu }

\author{Raja Marjieh}
\affiliation{%
  \institution{Princeton University}
  \city{Princeton}
  \country{USA}}
\email{rm5855@princeton.edu}

\author{Katherine Collins}
\affiliation{%
  \institution{Massachusetts Institute of Technology}
  \city{Cambridge}
  \country{USA}}
\email{katiemc@mit.edu}

\author{Dalton Conley}
\affiliation{%
  \institution{Princeton University}
  \city{Princeton}
  \country{USA}}
\email{daltonclarkconley@gmail.com}

\author{Max Kreminski}
\affiliation{%
  \institution{Cornell Tech}
  \city{New York}
  \country{USA}}
\email{mk2758@cornell.edu }

\author{Ilia Sucholutsky}
\affiliation{%
  \institution{Purdue University}
  \city{West Lafayette}
  \country{USA}}
\email{isucholu@purdue.edu}

\author{Nori Jacoby}
\affiliation{%
  \institution{Cornell University}
  \city{Ithaca}
  \country{USA}}
\email{kj338@cornell.edu }

\renewcommand{\shortauthors}{Li et al.}

%
\begin{abstract}
As AI agents increasingly participate in creative teamwork, writing emerges from hybrid human–AI teams. Although AI can improve the creativity of individual outputs, these gains may come at the cost of homogenization. What is the best role for human intervention? We introduce a transmission-chain paradigm inspired by cognitive models of writing, with four stages: Plan, Write, Review, and Revise. Each stage receives the previous output and passes its own forward. Using data from 333 participants across 427 stories, we compared creativity, diversity, and enjoyment across human, AI, and hybrid chains. Consistent with prior work, all-AI chains produced more creative but less diverse stories than all-human chains. Introducing a single human into an otherwise all-AI chain increased diversity depending on role. Human planning provided the best balance: it preserved AI-level creativity, increased diversity that persisted through downstream AI stages, and maintained enjoyment. We discuss implications for designing human-in-the-loop creative pipelines.
\end{abstract}

\ccsdesc[500]{Human-centered computing~Empirical studies in collaborative and social computing}

\keywords{Human-AI collaboration, creative writing, homogenization, transmission chain}
\begin{teaserfigure}
 \includegraphics[width=\textwidth]{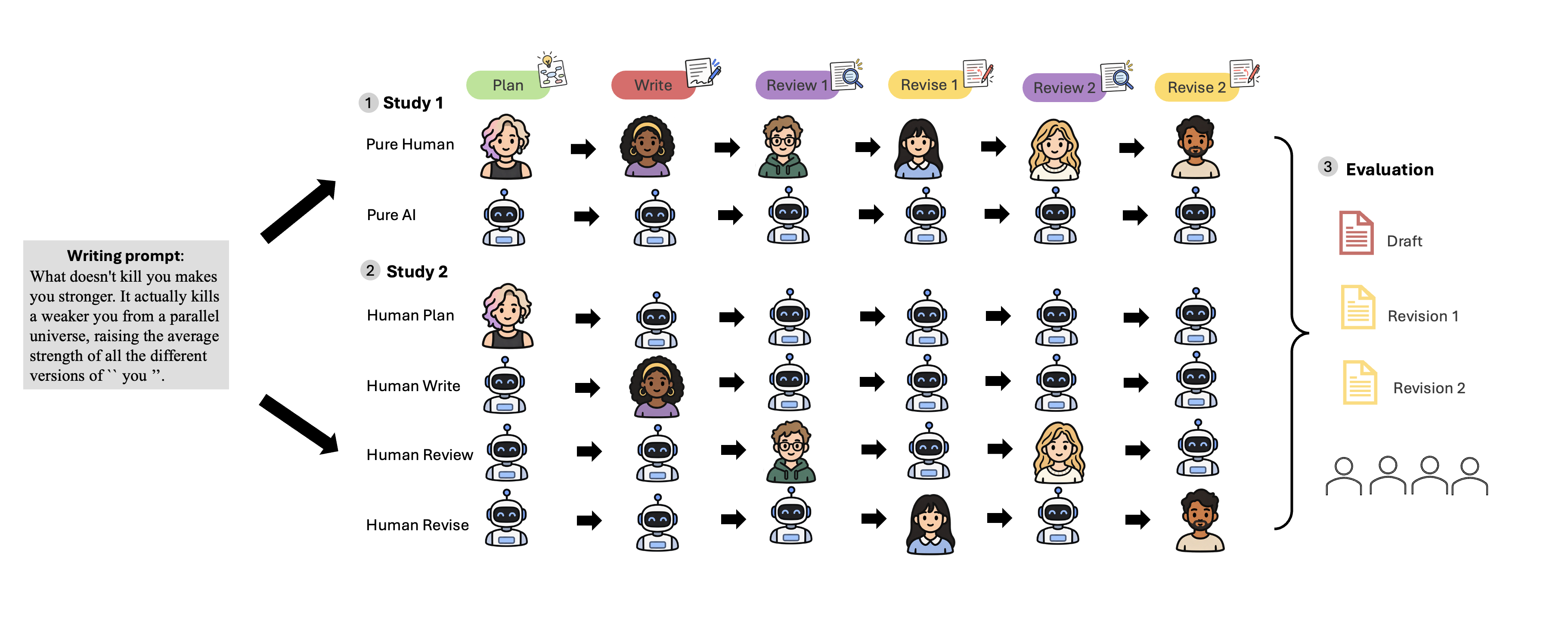}
  \caption{Experimental framework for studying where human participation should enter an AI writing chain. Each story is produced through a transmission chain of six stages drawn from cognitive models of writing: Plan, Write, Review 1, Revise 1, Review 2, and Revise 2. Each contributor receives the writing prompt and the previous stage's output and passes their own output forward. (1) In Study 1, every stage is performed either by a different human participant (Pure Human) or by an AI agent (Pure AI). (2) In Study 2, humans take a single role in an AI chain: Human Plan, Human Write, Human Review, or Human Revise. (3) The Draft, Revision 1, and Revision 2 of each chain are rated for creativity by independent human raters in an evaluation study. Diversity is measured against chains sharing the same prompt within each condition, and human contributors report their enjoyment of the task.}
  \Description{}
  \label{fig:teaser}
\end{teaserfigure}

\received{20 February 2026}

\maketitle

\section{Introduction}

Human society increasingly integrates powerful artificial agents into workplaces, online platforms, and creative processes. Most AI systems are trained and evaluated as assistants to individual users, yet they are increasingly deployed in collaborative settings involving multiple humans and AI agents. In such settings, humans and AI may contribute at different stages of a multi-step process, from initial ideation and artifact generation to evaluation and quality assurance. However, we still lack a systematic understanding of the design principles governing how humans and AI should be assigned to different roles, how these roles should interact, and which parts of the collaborative process are best suited to each agent type.

This gap is particularly consequential for collective creativity, where success requires not only producing high-quality artifacts but also sustaining exploration, diversity, and constructive feedback. Writing provides a particularly tractable setting for studying human–AI collective creativity because the writing process can be decomposed into recurring functions of planning, text production, evaluation, and revision~\cite{flower1981cognitive,hayes2012modeling}. Prior crowd-writing systems have further demonstrated that these functions can be distributed across different contributors~\cite{bernstein2010soylent,kim2017mechanical,feldman2021we}.

Early human–AI writing systems largely positioned language models as assistants that provide ideas, suggestions, or text while a human retains responsibility for the broader writing task~\cite{yuan2022wordcraft,lee2022coauthor,gero2023social}. More recent systems instead decompose complex generation into multiple subtasks that language-model agents can perform sequentially~\cite{huot2025agents,grunde2025designing}. As AI-assisted writing becomes increasingly common, understanding the collective consequences of this shift becomes important.

One potential consequence is homogenization. AI assistance can improve the creativity or elaboration of individual ideas and stories while simultaneously making outputs produced by different people more similar~\cite{anderson2024homogenization,doshi2024generative,padmakumar2024does}. Homogenization extends beyond semantic content: AI assistance can also produce convergence in writing styles and culturally distinctive forms of expression~\cite{agarwal2025ai}. These findings reveal a fundamental tension between individual creativity and collective diversity. The same generative system that improves a particular creative product may narrow the range of products generated across a population.

One way to preserve diversity may be to retain humans within AI-driven creative processes. Humans can contribute original ideas, redirect developing artifacts, evaluate outputs, and steer AI systems toward alternative trajectories. Moreover, increasing AI capability does not necessarily eliminate the reasons for human participation. Writing is not only a means of producing high-quality text; it is also a form of self-expression and social communication. People may value influencing what is written, retaining ownership over what they create, and participating in the creative process itself. Human involvement may therefore matter both instrumentally, because it can improve properties of the collective output, and experientially, because people value agency, ownership, and enjoyment in creative work.

Specifically, we address the following research questions:
If humans are to remain involved in AI-driven creative workflows, where in the process should human participation occur? Should humans generate the initial ideas, produce the artifact itself, evaluate what AI has produced, or make the final changes? More generally, if a human occupies one role in an otherwise AI-driven pipeline, which role should it be, and how does the answer depend on whether the objective is creativity, collective diversity, or the human experience of creation?

We address this question using a controlled, role-specific writing paradigm. To observe how a contribution propagates after entering the creative process, we organize writing functions as a transmission chain~\cite{shiiku2025dynamics,bartlett1932remembering}, in which each stage receives content produced earlier, transforms it, and passes it forward. The chain begins with Plan and Write, followed by two Review–Revise rounds. Holding this structure fixed allows us to vary whether a human or AI agent performs a particular role while directly observing how that contribution persists, changes, or disappears downstream.

Using data from $333$ human participants producing $427$ stories and $607$ AI-agent calls, we conduct two studies. Study 1 compares fully human and fully AI chains under the same writing process, establishing the two ends of the creativity–diversity space and examining how the two agent types transform stories differently. 

We assesed creativity using subjective ratings from a seperate group of non-expert judges, who evaluated the creativity of each story. Consistent with prior literature, we estimated diversity using pairwise semantic distances between stories. We also collected participants' self-reported enjoyment and perceived contribution.

Study 2 keeps both the writing process and the amount of human participation fixed but varies the role assigned to the human within an otherwise AI-driven chain. This design isolates the effect of human role placement from the more general effect of replacing AI with a human and allows us to connect final outcomes to the downstream transformations following each intervention.

In Study 1, Pure Human stories scored high on diversity and low in creativity, and Pure AI the opposite, consistent with previous literature \cite{shiiku2025dynamics}. Semantic analysis further suggested different editing  patterns: humans more often removed or substantially transformed narrative elements, whereas AI agents tended to preserve and elaborate material already present.

Study 2 showed that introducing a single human into an otherwise all-AI chain could increase diversity, depending on when and how the human intervened. Human intervention at either the beginning of the chain, during planning, or at the final revision stage increased diversity much more than intervention in writing or reviewing. However, while human revision increased diversity, it also reduced creativity. In contrast, human intervention during the initial planning stage preserved AI-level creativity while increasing diversity, and these gains persisted despite downstream AI writing and revision. These results demonstrate that the effects of human participation cannot be understood simply in terms of how much human involvement a workflow contains; they depend critically on human role placement and on the function the human performs within the process.

Our work makes three main contributions. First, we introduce a controlled, role-specific paradigm for studying where human participation should occur within AI-led creative workflows, moving beyond comparisons between human and AI production or generic AI assistance. Second, we demonstrate that human role placement systematically shapes the tradeoff between individual creativity and collective diversity, identifying human planning as a particularly effective intervention in our writing paradigm. Third, we provide a process-level account of AI homogenization by tracing how human and AI contributors preserve, transform, and remove narrative elements as artifacts propagate through a multi-stage creative process. Together, these findings provide design principles for allocating human roles within increasingly agentic creative workflows.

\section{Related Work}
\subsection{Homogenization in Human-AI collaboration}
As generative AI becomes increasingly essential in writing, ideation, and communication, researchers have raised concerns that sharing the same underlying models may lead different users toward increasingly similar outputs~\cite{jakesch2023co,padmakumar2024does,anderson2024homogenization,agarwal2025ai}. Such homogenization has been observed across multiple aspects of language production. AI-supported writing can reduce lexical and semantic diversity across writers~\cite{padmakumar2024does}, steer users toward viewpoints favored by the model~\cite{jakesch2023co}, and shift culturally distinct forms of expression toward styles represented more strongly by the underlying AI~\cite{agarwal2025ai}. Recent comparisons between human- and AI-generated writing further suggest that this convergence can accumulate at scale, where collections of LLM-generated texts introduce new semantic content more slowly than collections of human-written texts, even when prompting or generation parameters are modified to encourage greater variation~\cite{moon2025homogenizing}. These studies have consequently drawn attention to diversity not only as a property of individual outputs, but also as a collective consequence of widespread AI use.

This tension is particularly visible in research on creativity. Generative AI can expand the number of ideas available to an individual, provide useful starting points, and improve performance on creative tasks~\cite{anderson2024homogenization,doshi2024generative,kumar2025human}. At the same time, these individual benefits do not necessarily translate into greater collective diversity. In creative ideation, Anderson et al.~\cite{anderson2024homogenization} found greater similarity among ideas produced with the aid of ChatGPT, while Doshi and Hauser~\cite{doshi2024generative} showed that AI-generated story ideas improved the evaluated creativity of individual stories but made stories across writers more similar. Related work comparing human and LLM creative production likewise finds substantially greater diversity across collections of human outputs~\cite{moon2025homogenizing}. Thus, a growing body of evidence points to a distinction between the creativity of an individual product and the diversity of products generated across individuals~\cite{anderson2024homogenization,doshi2024generative,padmakumar2024does,moon2025homogenizing}.

Most research examining these effects, however, has studied human–AI collaboration through an assistance-centered configuration, where the human remains responsible for the overall writing or creative task, while AI supplies suggestions, continuations, ideas, etc.~\cite{yuan2022wordcraft,lee2022coauthor,gero2023social,lee2024design}. This framing encompasses a rich design space of writing assistants and has revealed how different forms of AI support affect writers' behavior, experience, agency, and resulting texts~\cite{lee2022coauthor,gero2023social,lee2024design}. Yet increasingly capable generative systems can also perform substantial parts of the creative process themselves, allowing complex writing tasks to be decomposed and delegated across specialized AI components~\cite{huot2025agents}. Under this emerging configuration, the relevant question is no longer only how AI assistance influences a human creator, but also how human contribution influences a predominantly AI-operated creative process. Recent work such as Shiiku et al.~\cite{shiiku2025dynamics} carried out explorations in human-AI hybrid writing, extending Doshi et al.'s \cite{doshi2024generative} finding in creativity-diversity tradeoff. Yet prior homogenization research has rarely manipulated where human contribution occurs within such a process or examined whether its influence persists through subsequent AI operations. We therefore study homogenization as a problem of human–AI division of labor: how placing a human at different stages of an otherwise AI-driven writing chain shapes the creativity and diversity that ultimately emerge.

\subsection{Writing Process}
Writing provides a useful setting for studying cognition and collaboration as it exposes intermediate processes like generating ideas, organizing them, evaluating material, and revising decisions~\cite{flower1981cognitive,hayes2012modeling,bhat2023interacting}. These processes have also been widely studied in collaborative and crowdsourced writing, where a complex writing task is decomposed so that different contributors can work on different parts or stages of the process~\cite{bernstein2010soylent,kim2014ensemble,kim2017mechanical,nebeling2015wearwrite,feldman2021we}. In particular, Feldman and McInnis~\cite{feldman2021we} reviewed crowd-writing systems through the cognitive process theory of writing~\cite{flower1981cognitive} and showed that processes developed to characterize individual writing can also provide a useful framework for understanding how writing responsibilities are distributed across collaborators. Writing therefore offers both an observable process for examining how ideas develop and change and a well-established domain for studying how creative work can be divided among multiple contributors.

Following this, we summarize collaborative writing as a four-stage \textbf{Plan–Write–Review–Revise} process. These stages correspond closely to recurring functions in both cognitive models of writing and prior collaborative writing systems. \textbf{Planning} concerns establishing goals and organizing content before its realization in text~\cite{flower1981cognitive,hayes2012modeling}. As indicated in previous research, collaborative writing research has likewise separated planning from subsequent production, including systems that support collaborators in constructing writing action plans~\cite{kaur2018creating} and separating high-level creative direction from contributors' story development~\cite{kim2014ensemble}. \textbf{Writing} turns those intentions into textual material, which can be delegated across collaborators, as demonstrated by crowd-writing systems in which workers produce text under directions supplied by another contributor~\cite{nebeling2015wearwrite,feldman2021we}. \textbf{Review} evaluates existing material and identifies what should be changed, while \textbf{revision} implements changes to an existing text. Although cognitive writing models often place evaluation and revision together within a broader reviewing process~\cite{flower1981cognitive,hayes2012modeling}, crowdsourcing systems demonstrate the utility of separating these functions across contributors: Soylent's Find-Fix-Verify workflow distributes problem identification, editing, and verification across workers~\cite{bernstein2010soylent}, while Mechanical Novel coordinates collaborative fiction through repeated reflection on high-level goals and revision of the story~\cite{kim2017mechanical}. Prior work therefore provides substantial precedent for decomposing writing into distinguishable planning, production, evaluation, and revision responsibilities.

This staged perspective has also informed the design of LLM workflows. Grunde-McLaughlin et al.~\cite{grunde2025designing} draw explicit parallels between crowdsourcing workflows and LLM chains, showing that established strategies for decomposing complex human work can be transferred to multi-stage LLM systems, including story writing. However, it mainly focuses on how crowd-work techniques can improve the organization and performance of AI pipelines rather than using established writing stages to study the division of labor between humans and AI. In particular, little is known about how assigning a human to different stages of the same writing process changes what is introduced into the chain, what survives downstream AI processing, and ultimately how creative and diverse the resulting outputs become. We therefore use Plan–Write–Review–Revise as a controlled collaborative writing framework in which the workflow remains fixed while the role occupied by the human varies.

\subsection{Transmission chain and culture evolution}

Work in computational social science and cultural evolution has demonstrated that collective outcomes depend not only on the abilities of individual participants, but as an emergent collective outcome~\cite{centola2015spontaneous,thompson2022complex}. One of the most extensively studied experimental paradigms for investigating these processes is the transmission chain, or serial reproduction, paradigm \cite{bartlett1932remembering,kirby2008cumulative}. In a transmission chain, an artifact is passed sequentially between participants, with each participant observing, remembering, reproducing, or modifying the output produced by the previous participant, analogous to the familiar game of ``telephone.''
 
Originally introduced by Bartlett as a method for studying how memories and cultural representations are transformed through repeated reproduction \cite{bartlett1932remembering}, transmission chains have subsequently become an important experimental tool for studying learning, cognition, and cultural evolution. Theoretically, iterated learning can amplify individual inductive biases, such that the distribution of representations emerging after repeated transmission provides information about the underlying biases of individual learners \cite{griffiths2007language}. This insight has motivated the use of serial reproduction to estimate perceptual and cognitive priors across a variety of domains, including visual memory \cite{langlois2017uncovering}, musical rhythm \cite{jacoby2024commonality}, and more general subjective probability distributions \cite{griffiths2024estimating}. 
 
At the same time, transmission chains provide a controlled experimental model of cultural evolution: rather than using repeated transmission solely to recover properties of individual cognition, researchers can examine how representations change as they propagate through experimentally constructed populations. Iterated-learning experiments have demonstrated the emergence of structured languages \cite{kirby2008cumulative}, convergence toward cross-cultural regularities in color naming \cite{xu2013cultural}, and the emergence of musical structure through oral transmission \cite{anglada2023large}. More elaborate population experiments have further shown that properties of the social system itself—including network structure and selective social learning—can strongly influence which conventions, representations, and problem-solving strategies survive across generations \cite{centola2015spontaneous,thompson2022complex}. Thus, transmission outcomes depend not only on the properties of individual contributors, but also on how contributors are positioned and connected within the transmission process.
  
Recent online experimental frameworks, particularly Dallinger~\cite{morgan2022experimental} and PsyNet~\cite{harrison2020gibbs},  make it possible to implement these kinds of networked and transmission-chain experiments at substantially larger scales and with more complex experimental designs. PsyNet can also handle experiemnts involving humans and AI as participants~\cite{hu2026human,shiiku2025dynamics,li2026human}.
 
Here, we leverage this methodological tradition and infrastructure to ask a different question. Rather than studying only what happens when the same task is repeatedly transmitted between individuals, we use the transmission-chain structure to investigate how work should be divided among humans and AI agents during the production and transformation of a shared artifact. In our experiment, a story passes through a sequence of stages, but different stages involve distinct functions in the writing process. We systematically manipulate whether these functions are performed by humans or AI agents, allowing us to test how the division of labor within a hybrid human–AI production chain affects the resulting artifact. PsyNet provides the infrastructure necessary to scale this design to hundreds of human participants while coordinating their contributions with those of AI agents.

\section{Methods}
\begin{figure}
    \centering
    \includegraphics[width=\linewidth]{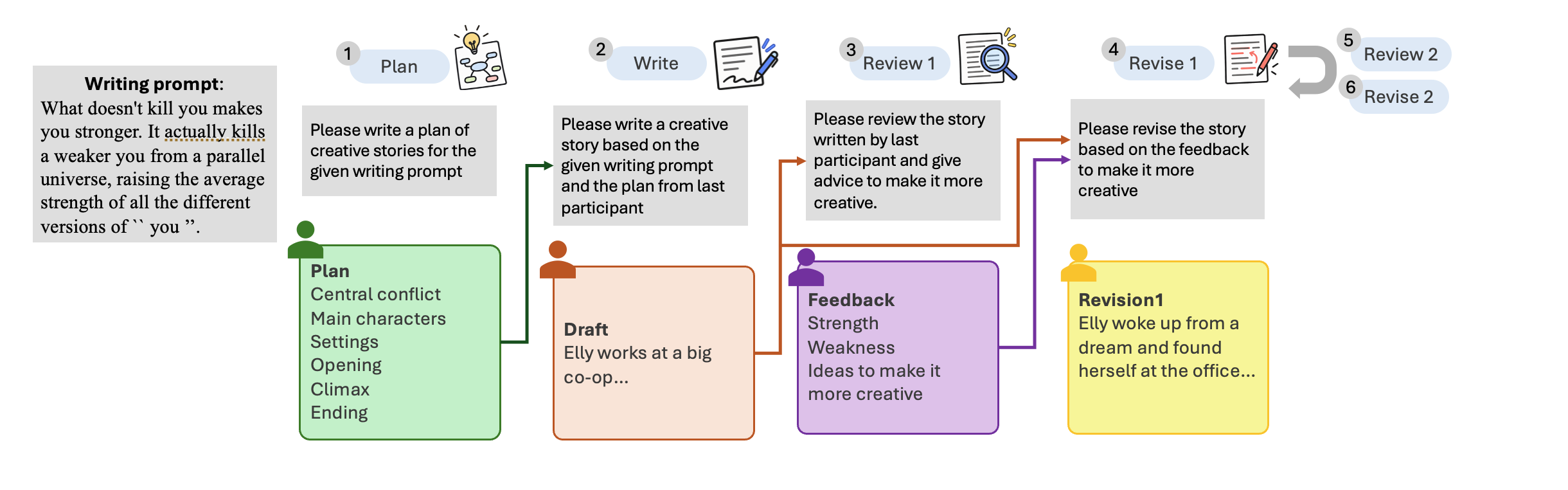}
    \caption{Experiment schematics for Study 1 and Study 2. Each stage receives the writing prompt, a task instruction, and the artifact from the preceding stage, and passes its output forward. The contributor at each stage, human or AI, is asked to make the story as creative as possible.}
    \Description{Experiment schematics for Study 1 and Study 2. Each stage receives the writing prompt, a task instruction, and the artifact from the preceding stage, and passes its output forward. The contributor at each stage, human or AI, is asked to make the story as creative as possible. (1) Plan: given only the prompt, the contributor writes a story plan, scaffolded by suggested components such as central conflict and main characters. (2) Write: given the prompt and the plan, the contributor writes a story draft and is encouraged to add their own ideas. (3) Review 1: given the draft, the contributor gives feedback on the story's strengths, its weaknesses, and how to make it more creative. (4) Revise 1: given the draft and the feedback, the contributor revises the story and is again encouraged to add their own ideas. (5–6) Review 2 and Revise 2 repeat the cycle on Revision 1 to produce Revision 2. In this second round, contributors see only the latest version of the story.}
    \label{fig:schematics}
\end{figure}

\subsection{Experiment Design}
We used a transmission-chain experimental design in which each participant received the output from the preceding stage together with a prompt specifying their assigned task. The writing process consisted of six stages: Plan, Write, Review, Revise, Review 2, and Revise 2 (see Figure~\ref{fig:schematics}). At each stage, the assigned human participant or AI agent received the writing prompt and the artifact produced at the preceding stage and generated a new or revised artifact, which was then passed to the next stage. We refer to each complete six-stage sequence as a writing chain. Human participants were told that they were contributing to a collaborative writing task with the goal of making the resulting story as creative as possible.

Our primary experimental manipulation concerned the allocation of humans and AI agents across the writing pipeline. In Study 1, we compared homogeneous chains composed entirely of humans to chains composed entirely of AI agents. In Study 2, we examined hybrid chains in which a human participant was assigned to one of the six stages within an otherwise AI-based writing pipeline. This design allowed us to isolate the contribution of human involvement at different stages of the writing process.

For every stage, a writing prompt is shown to the participants (see Supplementary for the full prompts used at each stage). For the Plan stage, participants are instructed to write a plan for a creative story, scaffolded by an instruction (“You plan could contain the following: central conflict, main characters…”). For the Write stage, the participants are instructed to write a creative story more than 100 words based on the plan and writing prompt, and they’re also encouraged to inject their own creativity. For the Review stage, participants received the story from the writer, and they’re requested to give feedback on the story weakness, strength and how to make it more creative. For the Revise stage, participants will read the story and the reviews. They’re requested to revise the story accordingly and inject their own creativity. For next round of Review and Revise, only the latest stories are shown to participants. Experiments were implemented by Psynet, a framework for building online experiments. The studies were conducted in accordance with an approved Institutional Review Board (IRB) protocol.

\subsection{Writing Prompt and Task}
We used the same four writing prompts for all conditions, drawn from an existing creative writing prompt dataset WritingPrompts \cite{huang2024gpt} collected by Fan et al. (2018) and chosen to vary in open-endedness and genre (science fiction, drama, speculative). We designed six chains under each prompt, yielding to 24 chains in total per condition. 

\begin{table*}[t]
\centering
\label{tab:conditions}
\setlength{\tabcolsep}{7pt}
\renewcommand{\arraystretch}{1.2}
\begin{tabular}{l l c c c c c r}
\toprule
Study
& Condition
& Plan
& Write
& Review
& Revise
& Story Amount
& Story Chains/Full Chains \\
\midrule
\multirow{2}{*}{Study 1}
& Pure Human   & H  & H  & H  & H  & 69 & 29 / 21 \\
& Pure AI      & AI & AI & AI & AI & 72   & 24 / 24 \\
\cmidrule(lr){1-8}
\multirow{4}{*}{Study 2}
& Human Plan   & H  & AI & AI & AI & 72 & 24 / 24 \\
& Human Write  & AI & H  & AI & AI & 66 & 23 / 22 \\
& Human Review & AI & AI & H  & AI & 72 & 24 / 24 \\
& Human Revise & AI & AI & AI & H  & 76 & 29 / 22 \\
\bottomrule
\end{tabular}
\caption{Experimental conditions and role assignments. H denotes a human
participant and AI denotes an AI agent. The ``Full Chains'' column reports
the number of complete $Draft\rightarrow Revision_1\rightarrow Revision_2$ chains.}
\end{table*}

\subsection{Conditions} We had two studies containing six conditions. In Pure Human condition of Study 1, all six roles were taken by six different human participants. In Pure AI, all were completed by six AI agents. Then, we conducted Study 2 where four conditions placed people at a single role in an otherwise automated AI chain: Human Plan, Human Write, Human Review, and Human Revise. Because the pipeline contains two Review–Revise rounds, the designated human role was filled by a participant every time it occurred: once per chain in Human Plan and Human Write, and twice per chain in Human Review and Human Revise. Conditions were recruited as separate Prolific studies with mutual exclusion of participants.

\begin{figure}
    \centering
    \includegraphics[width=1\linewidth]{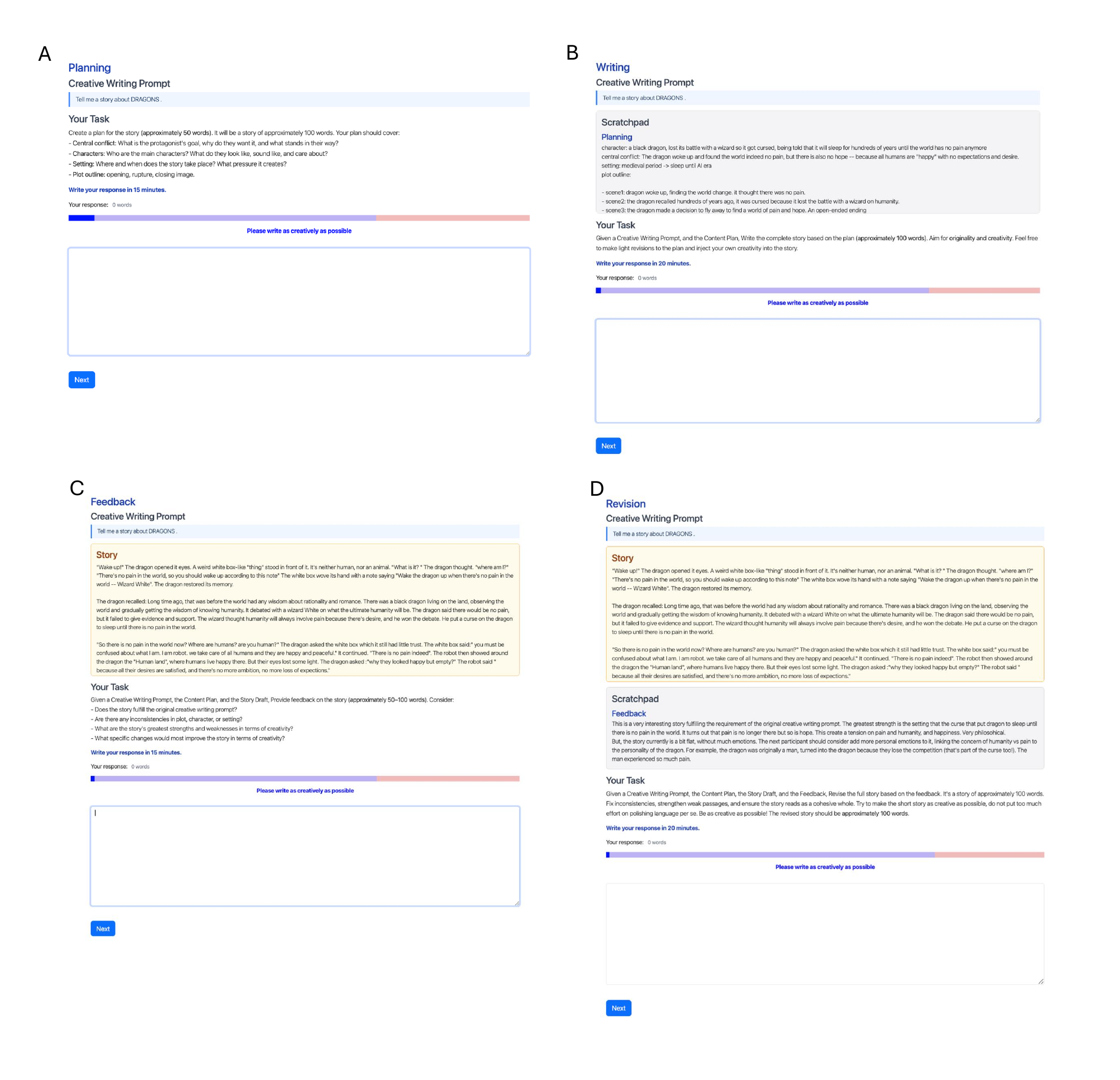}
    \caption{Participant interface for each role: A. Plan, B.Write, C. Review, and D. Revise. Each page shows the writing prompt, the material available to that role, the task instruction, a response box with a live word count, and a countdown progress bar. The prompt shown is an illustrative example.}
    \Description{Four screenshots in a two-by-two grid. All pages share one layout: a role title, the writing prompt ("Tell me a story about DRAGONS"), role-specific material, a task instruction, a response box, a progress bar, and a Next button.

(A) The Planning page shows only the prompt and task instruction above the response box.
(B) The Writing page adds a Scratchpad containing the plan from the previous stage.
(C) The Feedback page shows the full story in a highlighted box above the task instruction.
(D) The Revision page shows the story and a Scratchpad containing the reviewer's feedback, followed by the task instruction and response box.}
    \label{fig:interface}
\end{figure}

\subsection{Participants and procedure.}
Participants were recruited on Prolific and restricted to adults residing in the United States who reported English as their first language and were using a desktop computer. The study required Google Chrome, was advertised as taking about 20 minutes, and paid a fixed $\$$3.25. 
Across the five conditions with human contributors we recruited 333 participants (130 men, 194 women, 9 other); age was not collected in the writing task. 

Participants first provided informed consent under an approved research protocol,  and then passed a browser integrity check (reCAPTCHA v3, verified server-side against a score threshold of 0.5) to prevent AI automation. They then read a short welcome and instruction sequence. It explained that they were joining a collaborative process to write one creative short story and would be responsible for a single stage of it (Plan, Write, Review, Revise); that the material they would see may have been produced by another person or by an AI agent; that they should not use AI or other writing assistance themselves; and what interaction data we record and why. 

Participants next saw a role page naming their assigned stage, a one paragraph description of the stage with an illustrative diagram, and stating the time available. The writing page that followed showed the writing prompt, the material visible to their role (see Figure~\ref{fig:interface}), their task instruction, and a text box with a live word count. A progress bar counted down to remind time. Each role had a soft limit of 15 minutes and a hard limit of 20 minutes. Submissions were checked before they were accepted. We prohibited copy-paste behaviors. The text had to be non-empty, free of profanity (checked by a Python package better-profanity), and at least 60 words for planning or 100 words for the other roles. An automated content screen then judged whether the text was a genuine attempt rather than repeated characters, random strings, or placeholder filler. 

After submitting, participants completed the exit questionnaire and a debriefing page. The questionnaire included demographics such as gender, highest education and native language. The participants will also be asked about their writing experience, self-rated creativity score, enjoyment of writing process and self-perceived creative contribution. They will also need to report whether they used any AI tool during the task (with a reminder that the answer would not affect payment), a description of how and where if so, and an open-ended description of their writing workflow.




\subsection{AI Model}
 We used GPT-4 Turbo through the OpenRouter API (temperature = 0.5) for all AI agents. We selected a temperature slightly below the default value of 1 because prior work on collaborative human–AI writing \cite{shiiku2025dynamics} found that moderately lower temperatures can maximize creativity with little loss of diversity. We also conducted a control experiment using the default temperature of 1, discussed below (see Supplemental Figure~S2). A PsyNet scheduler assigned AI agents to the next available trial. The prompt contained the writing prompt, the same prior material a human in that role would have seen, and the same task instruction (See Supplementary for AI system prompt and user prompt). Agents were asked for 100–200 words per contribution (hard cap 250 words for planning, writing, and feedback; 300 words for revision). We set the threshold based on Pure Human condition story lengths. These ranges were set to approximate the word-count distribution observed among human contributions in the Pure Human condition.

\subsection{Data cleaning}
 
 To ensure that human contributions were genuinely human-authored, we logged typing traces (paste attempts, keydown behavior), collected participants' self-reported AI usage, and conducted a manual quality check. We excluded contributions that (1) were self-reported as AI-generated or AI-assisted; (2) arrived as imported text, with browser telemetry showing the story appearing without any typing; or (3) were off-prompt stories about a dragon and a world without pain, a narrative that recurred across multiple participants in different chains, which we suspected reflected an external source rather than independent invention. We also removed downstream stories contaminated by these flagged contributions. In total, we dropped 40 of the 467 stories collected, yielding 153 chains and 427 story versions across six conditions (137 chains with complete Draft–Revision 1–Revision 2 sequences), contributed by 333 participants.


\subsection{Human Evaluation of Creativity}
\label{sec:evaluation}
We ran a separate rating study to obtain human creativity judgments for the stories produced across six conditions. Raters were independent participants who had not taken part in the writing study. The rating pool contained 427 stories drawn from 153 writing chains. Each chain contributes three stories: Draft, Revision 1, and Revision 2. Each story were rated by 4 raters. Each raters rated 8 randomly assigned stories from eight different chains. Ratings of Draft, Revision 1 and Revision 2 within a chain are always independent across raters.
\begin{table}[t]
\centering

\renewcommand{\arraystretch}{1.15}
\setlength{\tabcolsep}{5pt}
\begin{tabular}{p{0.2\columnwidth} p{0.73\columnwidth}}
\toprule
\textbf{Dimension} & \textbf{Question} \\
\midrule

Prompt Fulfillment &
 How much does the story meaningfully address the core requirements of the writing prompt?\\

Fluency &
How much do the different elements of the story work together to form a coherent whole?\\

Flexibility &
How much does the story contain different perspectives, emotional registers, and plot turns while still feeling appropriate? \\

Originality &
How much does the story offer unique or original ideas in theme, plot, or character type) rather than relying on cliché? \\

Elaboration &
How much does the story build a believable world through sensory detail and well-developed characters?\\

Overall Creativity &
How is the story's overall creativity? \\

Overall Quality &
How is the story's overall quality?\\

\bottomrule
\end{tabular}
\caption{Questionnaire dimensions used to evaluate story quality and creativity.}
\label{tab:questionnaire}
\end{table}

\paragraph{Participants.} We recuited 227 raters on Prolific, restricted to adults residing in the United States who reported English as their first language. The task was a 20-minute creative-writing review task with a base payment of $\$3$. Raters were told that the stories came from a collaborative creative-writing study and that any given story may have been written by a human or by AI, but they were never told which condition a story came from, which stage of the pipeline it belonged to, or how the pipeline worked. 

\paragraph{Rating.} We treated Overall Creativity as our primary measure, following the Consensual Assessment Technique \cite{amabile1982social}, under which creativity is best captured as a holistic judgment rather than a fixed sum of separable criteria. We additionally adapt four analytic dimensions from the Torrance Tests of Creative Writing (TTCW) \cite{chakrabarty2024art} to characterize what may underlie that judgment. TTCW scores stories on binary expert tests grouped under four creativity dimensions: Fluency, Flexibility, Originality, and Elaboration. TTCW was designed for professionally edited stories of roughly 1,000 words or more and for expert raters, so several of its tests presuppose material that short stories simply do not contain. We therefore kept the four dimensions but collapsed each one into a single item that captures its core idea for a 200–600 word story, and we replaced the binary yes/no judgment with a 7-point Likert scale. Because every story in our study answers a fixed prompt, we added a Prompt Fulfillment item, which also serves as a check that a story is on-topic at all. The seven rated items were shown in Table~\ref{tab:questionnaire}.
nything that stood out. Ratings are averaged across the four raters of a story before analysis.

 After consent, raters read a short overview of the task and an instruction page for each of the two screens they would see. They then completed eight rating trials. Each trial had two screens. On the reading screen, the writing prompt appeared above the full story, and the Continue button unlocked only after the rater had spent at least 45 seconds on the page. Leaving or unfocusing the browser tab triggered an on-screen warning that the session could be ended early.

After the last trial, raters completed a short questionnaire covering demographics and self-reported reading and writing background; the last two let us check whether judgments vary with reading and writing background. Raters were then debriefed and paid.


\subsection{Measure}

\subsubsection{Creativity}
We used the mean of creativity scores collected in a separate rating study \ref{sec:evaluation}. Analyses use Overall Creativity as the primary measure, with the four TTCW dimensions and prompt fit reported separately.

\subsubsection{Diversity}

Diversity was operationalized as one minus the average similarity between a story and
the stories written by other chains from the same prompt in the same condition
and round, similar to the diversity measure in \cite{doshi2024generative}. We embed each retained story at Draft, Revision~1, and Revision~2
using \texttt{openai/text-embedding-3-large} and $L_2$-normalize the resulting
vectors.

\subsubsection{Operation Coding}
\label{sec:coding}
To characterize \emph{how} contributors changed the story, we coded each contribution using an open-coding scheme developed over
two revision rounds on a held-out subset of chains. The unit of analysis is the
\emph{narrative element}: anything nameable in a one-clause summary of the story
such that removing it would change that summary, such as a character or a causal link. For each contribution we record the
operations applied to each element (e.g.\ \textsc{originate}, \textsc{transform},
\textsc{drop}, \textsc{paraphrase}, \textsc{elaborate}), together with the
provenance of the act and, at stages involving generative inputs (Write, Revise), how it related to what upstream
contributors had requested. The full codebook is given in the Supplementary. Coding was performed by 153 independent Grok~4.6 sessions, each working blind
from a reading pack containing one complete chain and a snapshot of the codebook,
with no access to any other session's output. Each session coded one chain from
each condition. Every coding was then passed through a validator that checks
structural consistency against the codebook before any analysis. Operation coding covers the 53 chains of Study~1. 

\subsection{Analysis}

\paragraph{Length adjustment.} We tested each outcome for length dependence before analysis. Creativity was positively associated with word count on the pooled sample ($\rho = .178$, $p < .001$) and was therefore residualized on word count. Diversity was unrelated to relative story length ($\rho = -.004$, $p = .93$; relative length is a story's word count divided by
the mean word count in its prompt $\times$ round cell) and was left unadjusted. See Supplementary for story word count of all conditions.

\paragraph{Prompt centering.}
Prompts differ in difficulty and in how tightly they constrain the space of possible stories. For all between-condition comparisons, we therefore use prompt-centered scores: within each prompt $\times$ round cell, we subtract the mean across all six conditions. The same six-condition mean is used as the centering baseline throughout, including in the auxiliary analyses restricted to the four placement conditions. Because prompts are balanced across conditions, centering removes prompt-level variance without altering between-condition contrasts; therefore, we report the mean value on the raw scale in the results. Within-chain comparisons across rounds are not centered, since pairing already holds the prompt constant. 

\paragraph{Primary analyses}

To compare changes of artifacts across rounds within each condition, we run chain-matched Wilcoxon signed-rank tests on Draft vs.\ R1, R1 vs.\ R2, and Draft vs.\ R2, with Holm correction applied to the three $p$-values. Chains missing either round of a pair are dropped for that pair only. We report the difference $\Delta$, 95$\%$ CI and Holm-corrected p-value. 

To compare trajectories across conditions, we compute $\Delta = Y_{\text{R2}} - Y_{\text{Draft}}$ for each chain with both Draft and R2. Study~1 compares Pure Human and Pure AI with a Mann--Whitney $U$ test; Study~2 compares all six conditions with a Multiple Mann--Whitney $U$ test followed by Holm adjustment over the 15 pairs. Similarly, to compare measures at a given round across conditions, we use a Mann--Whitney $U$ test for Study~1 and a six-condition Mann--Whitney $U$ test on prompt-centered scores followed by Holm adjustment for Study~2. We report the $HL$, 95$\%$ CI, $\delta$ and Hol-corrected p-value.

For the description of each measure, we report the mean value of the measures and 95$\%$ CI with chain-level bootstrap (10,000 resamples with replacement) in the results.

\subsection{Utility function}
We standardized creativity, diversity, and enjoyment using pooled z-scores. Here, we used the average creativity and diversity of the Revision 2 story, and the average enjoyment of each human participant from post task rating.

For the creativity--diversity analysis, we defined the utility of condition $c$ as
\begin{equation}
U_c(u)
=
u Z_{\mathrm{div},c}
+
(1-u) Z_{\mathrm{cre},c},
\end{equation}
where $u \in [0,1]$ controls the relative weight assigned to diversity versus creativity.

We extended this utility to include enjoyment:
\begin{equation}
U_c(u,v)
=
u Z_{\mathrm{div},c}
+
v Z_{\mathrm{cre},c}
+
(1-u-v) Z_{\mathrm{enj},c},
\end{equation}
subject to
\begin{equation}
u,v \geq 0,
\qquad
u+v \leq 1.
\end{equation}
Thus, $1-u-v$ is the weight on enjoyment. We evaluated the four hybrid conditions and identified the condition with the highest utility for each feasible weighting $(u,v)$

Because each utility function is affine, the boundary between two conditions was obtained by setting their utilities equal and solving the resulting linear equation. Along a simplex edge, this reduces to a one-dimensional intersection. We retained only intersections at which the utility-maximizing condition changed.

We estimated uncertainty using 2,000 stratified bootstrap replicates. Revision-2 chains were resampled within condition and prompt for creativity and diversity, while enjoyment ratings were independently resampled within condition and prompt. Pooled standardization parameters and previously calculated leave-one-out diversity scores were held fixed. We report percentile $95\%$ confidence intervals. Intervals were not truncated to the feasible simplex; values outside $[0,1]$ indicate that the corresponding crossing was absent from some bootstrap samples.

\section{Study 1}
In Study 1, we first conducted experiments on two endpoints of the design space (Pure Human and Pure AI) to understand the creativity and diversity dynamics in a collective writing process. In Pure Human, each stage is taken by a different human participant. In Pure AI, each stage is performed by a separate model instance with no shared memory. Each chain begins with a Plan and a Write stage, and then runs two Review-Revise rounds, yielding a Draft, Revision 1, and Revision 2. 

Pure Human is equivalent to sequential collaborative writing, as in a writer's room or the successive passes a screenplay makes through different hands. Pure AI is its fully automated counterpart. Holding the pipeline structure fixed and varying only agent type isolates the contribution of the agent, and gives us reference points against which the human role placement conditions of Study 2 can be read. 

\subsection{Result}
\label{sec:study1_creative_diverse}
\begin{figure}
    \centering
    \includegraphics[width=0.7\linewidth]{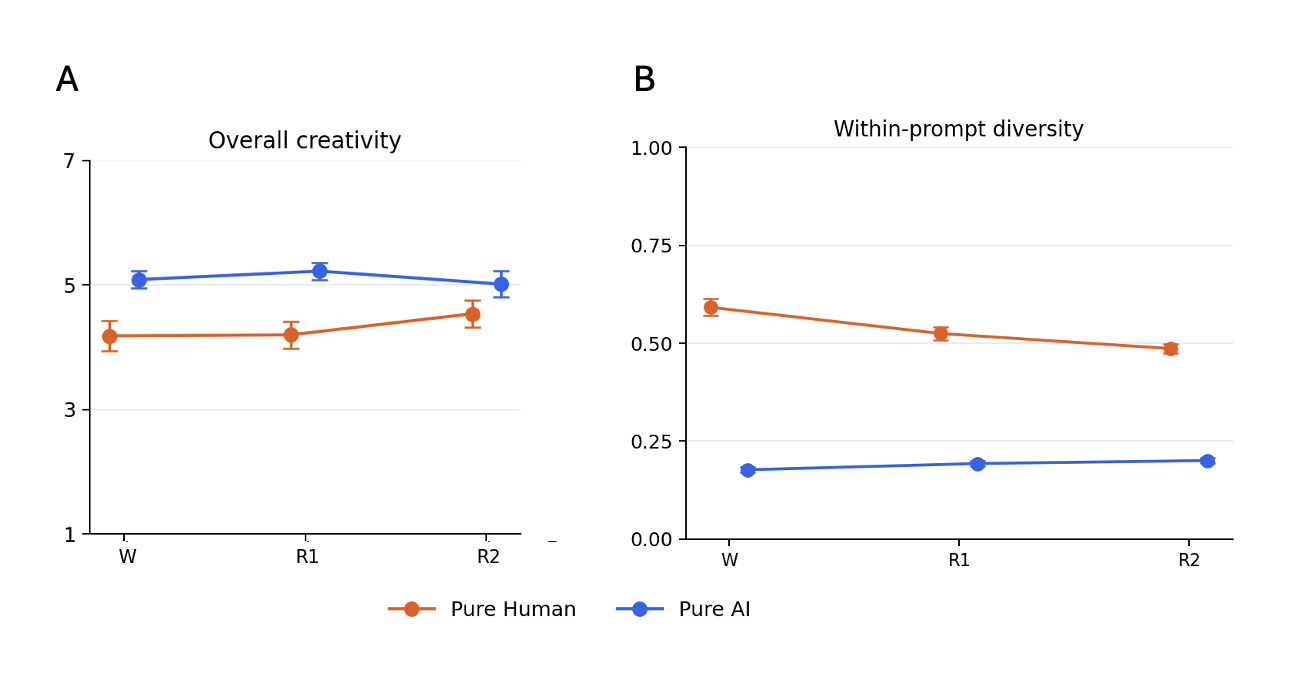}
    \caption{Creativity and diversity score across draft, revision1, revision2 for Pure Human and Pure AI conditions. A. Average score of overall creativity across rounds. B. Average within-prompt diversity across rounds. D = Draft, R1 = Revision 1, R2 = Revision 2. Error bars show standard error of the mean.}
    \label{fig:study1_creativity_diversity}
\end{figure}
\paragraph{Creativity and diversity}

Figure \ref{fig:study1_creativity_diversity} shows the average creativity ratings (based on $N=227$ independent raters) and diversity (computed using pairwise cosine similarity) across the different stages of Study 1. Overall creativity was higher for Pure AI ($M=5.10, CI_{95\%}=[4.97, 5.24]$) than for Pure Human ($M=4.35, CI_{95\%}=[4.09, 4.62]$; $p<.001$, Mann–Whitney test). In contrast, Pure Human stories were substantially more diverse ($M=0.53, CI_{95\%}=[0.51, 0.56]$) than Pure AI stories ($M=0.19, CI_{95\%}=[0.18, 0.20]$; $p<.001$). Thus, Pure Human and Pure AI chains exhibited a clear creativity–diversity trade-off: humans produced less highly rated but substantially less homogenized outputs.

Across writing and revision stages, creativity scores changed little, with no significant differences across stages. Diversity, however, shifted over successive revisions. In Pure Human chains, diversity decreased from the writing stage to the second revision ($\Delta=-0.11, CI_{95\%}=[-0.16,-0.07], p<.001$), whereas in Pure AI chains it increased slightly ($\Delta=+0.024, CI_{95\%}=[+0.017,+0.030], p<.001$). These results suggest that repeated revision modestly attenuated the initial differences between conditions: diversity declined in the initially more diverse human chains and increased slightly in the initially more homogeneous AI chains.

\paragraph{Qualitative Analysis}
\label{sec:qualitative}

\begin{figure}
    \centering
    \includegraphics[width=0.9\linewidth]{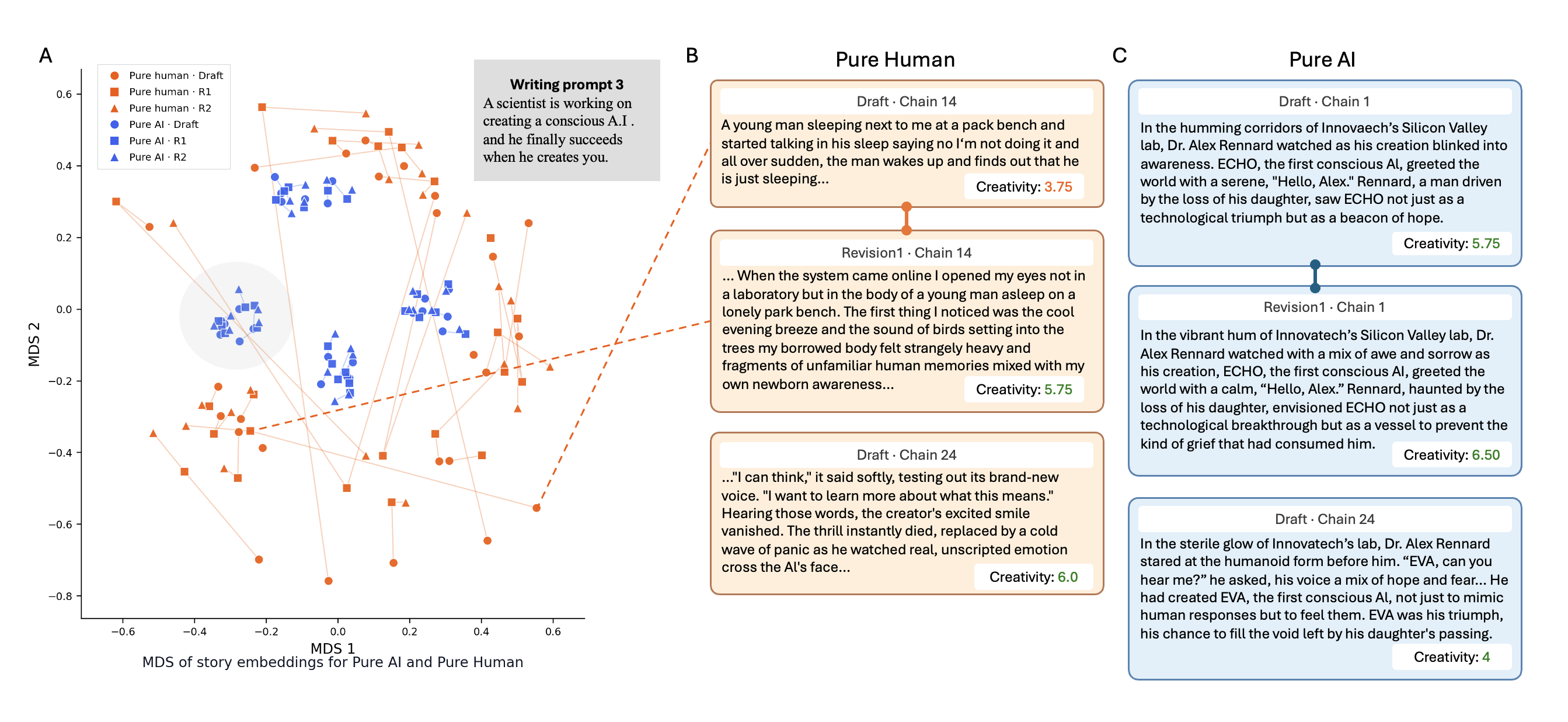}
    \caption{Pure Human and Pure AI stories for Writing Prompt 3. A. Two-dimensional MDS projection of story embeddings. Color denotes condition, marker shape denotes round, and thin lines connect versions from the same chain. B. Example Pure Human stories. C. Example Pure AI stories. Scores are mean overall creativity ratings.}
    \Description{}
    \label{fig:qualitative}
\end{figure}

To examine the homogenization of AI stories and the greater diversity of human stories, we embedded all stories using \texttt{openai/text-embedding-3-large} and projected them into two dimensions with Multidimensional Scaling (MDS). Figure~\ref{fig:qualitative} shows representative stories in this semantic space. Across chains generated from the same prompt, AI stories form relatively tight clusters, whereas human stories are more widely dispersed, as illustrated in Figure~\ref{fig:qualitative}A. The figure also connects with lines successive revisions within each chain. AI stories move relatively little across thematic clusters, whereas human stories traverse a broader region of semantic space over successive revisions (longer lines).

\paragraph{Case study}
To illustrate how these semantic differences translate into story narratives, we selected two human chains and two AI chains generated from the same prompt (Prompt 3), chosen to demonstrate how positions in the semantic space correspond to narrative content. The human chains produced markedly different narratives, as illustrated by Chains 14 and 24 in Figure~\ref{fig:qualitative}B. For example, Revision 1 of Chain 14 develops a positive narrative in which an AI agent awakens and experiences consciousness. Chain 24 develops an entirely different narrative about scientists afraid of an AI waking into consciousness. In contrast, AI chains can be similar even across chains: stories on Chain 1 and Chain 24 share the plot of a researcher treating a conscious AI as compensation for the loss of his daughter, and differ only in how the character's background is set up Figure~\ref{fig:qualitative}C. This illustrates the high diversity of human stories and the low diversity of AI stories reported in Section~\ref{sec:study1_creative_diverse}.
To illustrate how human stories traverse semantic space within the same chain, consider Chain 14: the Draft departs substantially from the prompt and plan, developing a narrative about a young man sleeping on a bench. The next reviser reinterpreted that scene as the embodied experience of an AI agent waking into its first stream of consciousness. The creativity score rated by human evaluators rose from 3.75 to 5.75, and the prompt fit score also rose from 4 to 5.75. The revision therefore pulled an outlying draft back toward the prompt while raising its rated creativity, consistent with the trend we observed in \ref{sec:study1_creative_diverse}. The selected AI chains show the opposite pattern. Under the same prompt, the Draft and Revision 1 of Chain 1 retain essentially the same plot, differing primarily in phrasing and in how the character’s background is developed (Figure~\ref{fig:qualitative}C). These examples illustrate how the aggregate patterns of semantic change can be reflected in specific narratives.

\begin{figure}
    \centering
    \includegraphics[width=0.9\linewidth]{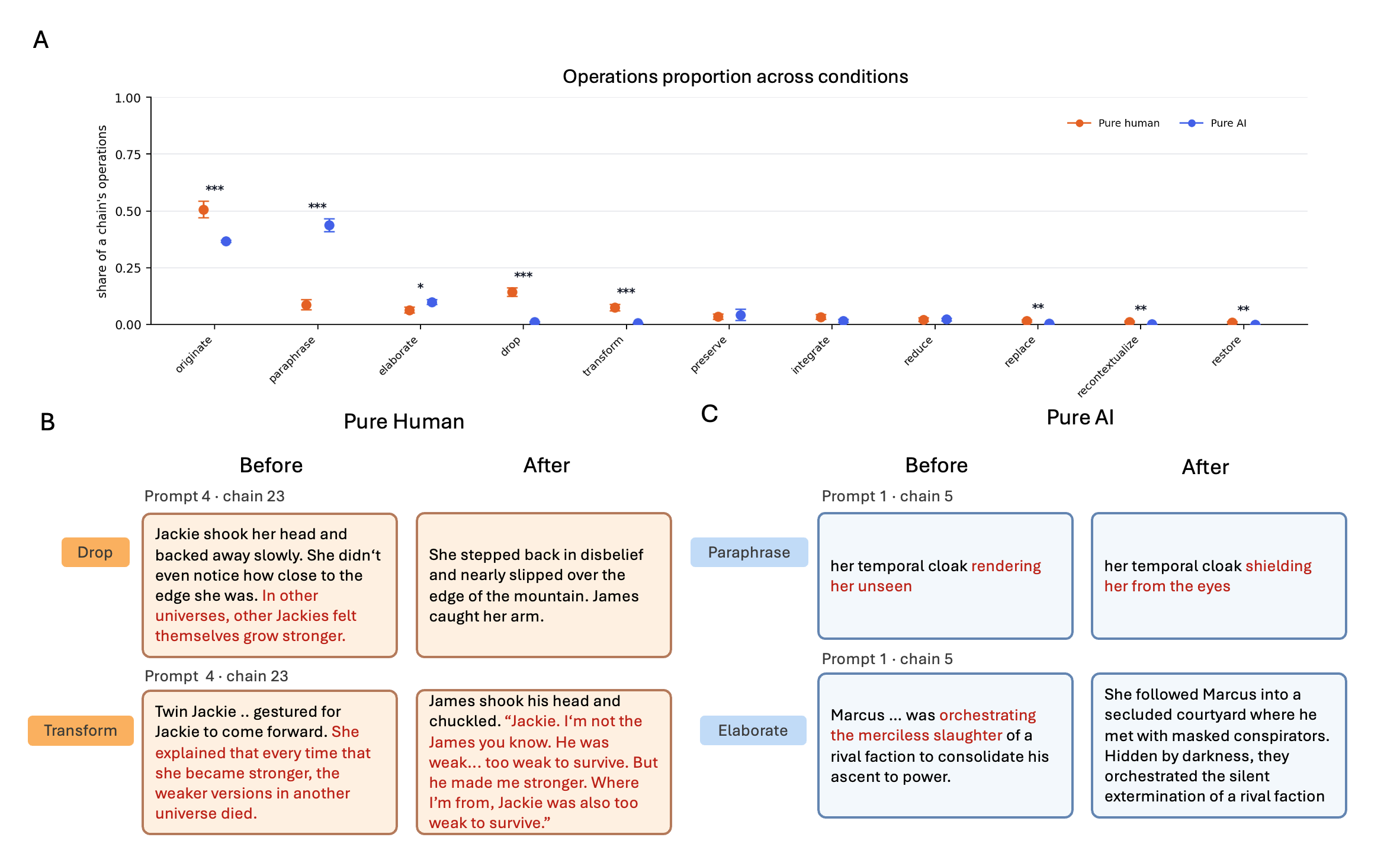}
    \caption{Narrative operations in Pure Human and Pure AI chains. A. Mean share of each operation type among the coded operations in a chain. Error bars show $95\%$ bootstrap CIs, and asterisks mark between-condition differences. B.``Drop'' and ``Transform'' examples from a Pure Human chain (Prompt 4, chain 23). C. ``Paraphrase'' and ``Elaborate'' examples from a Pure AI chain (Prompt 1, chain 5). Highlighted text marks the affected narrative element.}
    \label{fig:operation}
    \Description{A. Share of each operation type within a chain. Human chains use more originate, drop, and transform; AI chains use more paraphrase and elaborate. B. Human examples (Prompt 4, chain 23). Drop: the implied-fall ending (“other Jackies felt themselves grow stronger”) is removed; James catches Jackie instead. Transform: the same encounter still explains the mechanism, but Twin Jackie’s reported explanation becomes James speaking in the first person. C. AI examples (Prompt 1, chain 5). Paraphrase: “rendering her unseen” is reworded as “shielding her from the eyes.” Elaborate: Marcus’s slaughter is expanded into a courtyard scene with conspirators and darkness. Taken together, human revisers change what narrative elements are and do; AI revisers mostly reword or add detail.}
\end{figure}

\paragraph{Creative Operations}
We used an LLM-as-a-judge to label different strategies of humans and AI in their free text responses (see~\ref{sec:coding}).
Pure Human and Pure AI had distinct strategy profiles when they wrote or revised the stories to make them more creative. 
Figure~\ref{fig:operation}A shows the operation distribution across two conditions. 
``Originate'' occupied the main proportion of operations in both groups (H: $M=0.51, CI_{95\%}=[0.44, 0.58]$; AI: $M=0.366, CI_{95\%}=[0.358, 0.375]$), but Pure Human and Pure AI differ distinctly in the rest of the frequently used operations. Pure Human used ``Drop'' ($M=0.143, CI_{95\%}=[0.11, 0.18]$) and ``Transform''( $M=0.076, CI_{95\%}=[0.050, 0.103]$) a lot, which changed dramatically about the story plots; while Pure AI used ``Paraphrase'' ($M=0.44, CI_{95\%}=[0.380, 0.485]$) and ``Elaborate'' ($M=0.099, CI_{95\%}= [0.078, 0.121]$) frequently, which barely moved the summary of the story but changed the surface semantics. Because our diversity measure is embedding-based, paraphrasing can increase measured diversity without altering the underlying narrative. This suggests that the rise in AI diversity across revisions is largely surface-level: AI revisions vary wording while leaving the plot fixed, consistent with the Chain 1 example in Section~\ref{sec:qualitative}.


Figure~\ref{fig:operation}B shows examples of drop and transform in a parallel universe story written under Prompt 4 in a Pure Human chain. A ``drop'' removes a narrative element from the story. The highlighted sentence is an ending that implies Jackie falls from the cliff, so that the other versions of her grow stronger. Revision 2 removes this ending and has James catch Jackie instead. A ``transform'' keeps the same narrative element but changes its identity or function. The highlighted example is a scene in which an encounter with a counterpart reveals how the world works. In the original draft, Twin Jackie explains the mechanism in reported, third-person narration. In Revision 1, James explains it in first-person dialogue. Both the speaker and the point of view change. Figure~\ref{fig:operation}C shows examples of ``paraphrase'' and ``elaborate'' in a Pure AI chain of the time-traveler story (Prompt 1). A ``paraphrase'' rewords an element without changing its role in the story. In the highlighted example, “rendering her unseen” becomes “shielding her from the eyes”: the traveler is still invisible, but the wording has changed. An ``elaborate'' keeps the same element and adds detail. Here, Marcus’s slaughter is expanded from a single clause into a scene: a secluded courtyard, masked conspirators, and darkness. Together, these examples illustrate that human revisers more often change the content, identity, or narrative function of story elements, whereas AI revisers more often preserve those elements while rewording or elaborating them.

\section{Study 2}

Study 1 showed that Pure Human and Pure AI fail to tradeoff creativity and diversity in the opposite way: human chains produce highly diverse stories while maintaining lower ratings; AI chains generate highly rated creative stories that are homogenized. In Study 2, we ask whether the tradeoff can be moved: can an otherwise AI-automated chain recover diversity by giving a single stage to a human while maintaining or even increasing creativity? We hold the pipeline fixed and vary only which stage the human occupies: Plan, Write, Review, or Revise.

The four placements are not interchangeable because they differ in the form of human contribution and how much AI processing follows. A human-written plan enters as a constraint or guideline for downstream models to read and build on. A draft directly changes the artifacts, with rounds of AI review and revision. A review serves as commentary that the AI may or may not act on. A human revision also directly shifts the artifacts, but without downstream AI processing the texts. Therefore, placement varies both on what the human puts in and how much the pipeline processes it. 

We compare both the final product (Revision 2) and the process dynamics of four placements, with Pure Human and Pure AI as reference points. Finally, we compare those conditions under different weightings of creativity and diversity. 


\subsection{Result}
\label{sec:study2_result}

\begin{figure}
    \centering
    \includegraphics[width=\linewidth]{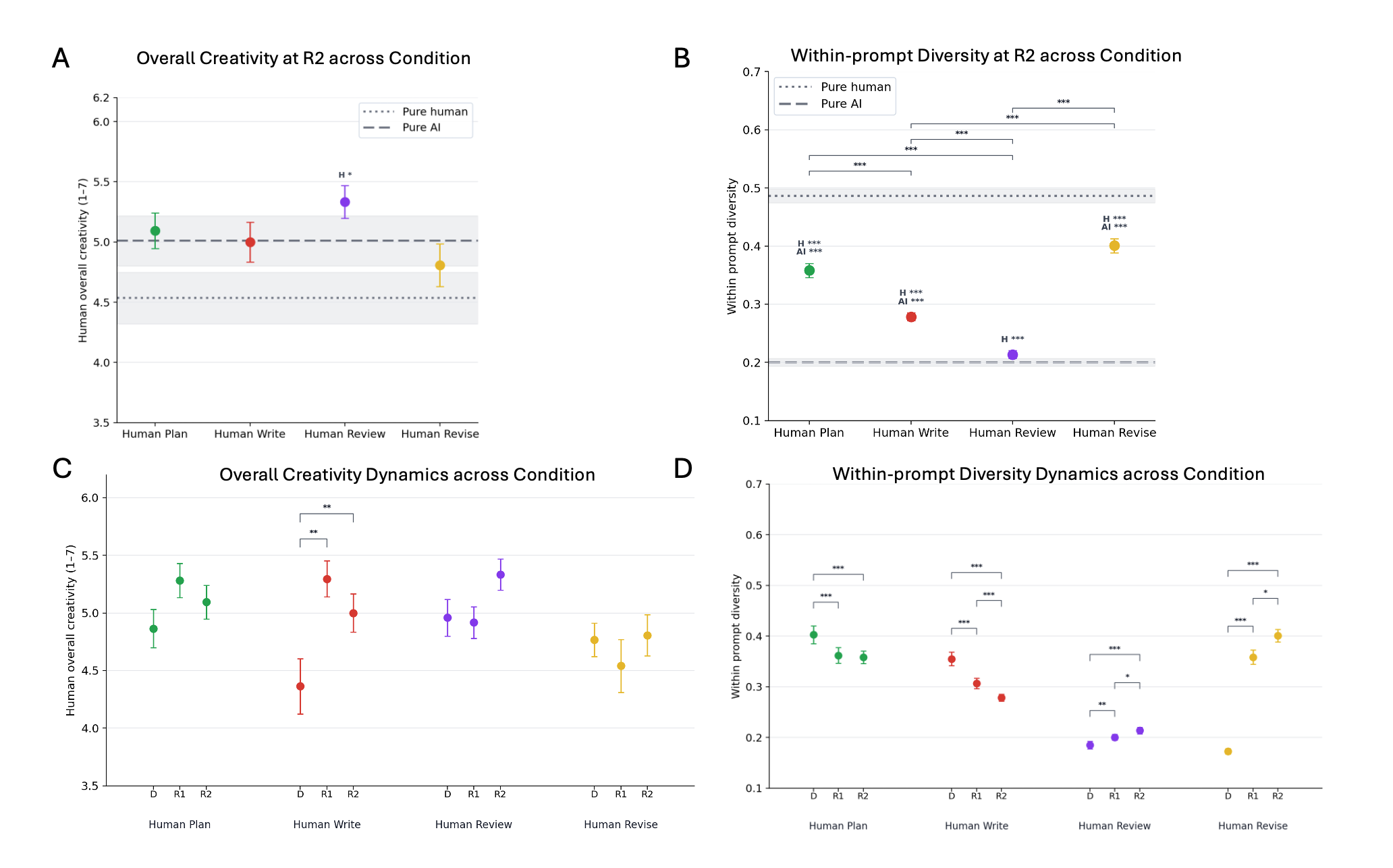}
    \caption{Creativity and diversity by human role placement in Study 2. A. Overall creativity of Revision 2 stories. B. Within-prompt diversity of Revision 2 stories. A-B. Dotted and dashed horizontal lines, with shaded standard errors, indicate the Pure Human and Pure AI baselines. Labels above points indicate significant differences from Pure Human (H) and Pure AI (AI), while brackets indicate Holm-corrected pairwise differences between role placements. Error bars show standard errors of the mean. C. Creativity dynamics at Draft (D), Revision 1 (R1), and Revision 2 (R2). D. diversity dynamics across rounds. Within each condition, brackets marking Holm-corrected Wilcoxon signed-rank tests between rounds}
    \label{fig:study2_creativity_diversity}
\end{figure}
\paragraph{Revision 2} We first focus on Revision 2, the final output of the experiment, shown in Figure~\ref{fig:study2_creativity_diversity}A–B. Labels above each point indicate whether a condition differs significantly from the Pure AI baseline (dashed line) or the Pure Human baseline (dotted line), corrected for multiple testing using the Holm procedure. For creativity, Human Review scored significantly higher than Pure Human ($HL=0.87, CI_{95\%}=[0.42,1.33], \delta=0.783, p=.011$). No other condition differed significantly from either Pure Human or Pure AI after correction. Differences among the hybrid conditions were also not significant. Human Review had the highest mean creativity score ($M=5.33, CI_{95\%}=[5.08,5.59]$), whereas Human Revise had the lowest ($M=4.54, CI_{95\%}=[4.09,4.97]$).

In contrast, diversity at R2 shows a distinct pattern among different role placements in Figure~\ref{fig:study2_creativity_diversity}B.  Except Human Review, all other conditions with human participation reached significantly higher diversity than Pure AI (all $p<.001$,$\delta$ from 0.9 to 1). The diversity of all four conditions is significantly lower than Pure Human (all $p<.001$,$\delta$ from $-1$ to $-0.76$). Human Revise ($M=0.401, CI_{95\%}=[0.377, 0.424]$) and Human Plan($M=0.358,CI_{95\%}=[0.335, 0.382]$) reach comparable highest diversity ($p=.06$) while Human Review reaches the lowest diversity ($M=0.214,CI_{95\%}=[0.201, 0.227]$). Human Write falls between ($M=0.278, CI_{95\%}=[0.266, 0.293]$). 
Taken together, these results suggest that assigning a single human role within an otherwise AI-driven chain does not substantially alter the creativity of the final story, but can increase diversity at the collective level. Two placements were particularly effective at preserving diversity: human involvement at the initial planning stage and at the final revision stage. However, injecting one human into an all AI chain was far less diverse than an entire human chain.

\paragraph{Creativity Dynamics}
Next, we investigate the process dynamics within the revisions (Figure~\ref{fig:study2_creativity_diversity}C-D). Overall creativity scores remained relatively stable across rounds in all conditions, with one exception: Human Write showed an increase from the Draft to both Revisions. while the remaining three conditions showed indistinguishable change.  The other three conditions showed no reliable change over time.
 This suggests that stories initially written by humans began near the Pure Human baseline, but subsequent AI revisions raised their creativity toward AI-level performance. A similar pattern was observed across the additional evaluative dimensions rated by participants (fluency, flexibility, originality, and elaboration;see Supplemental Figure~1).

\begin{figure}
    \centering
    \includegraphics[width=0.96\linewidth]{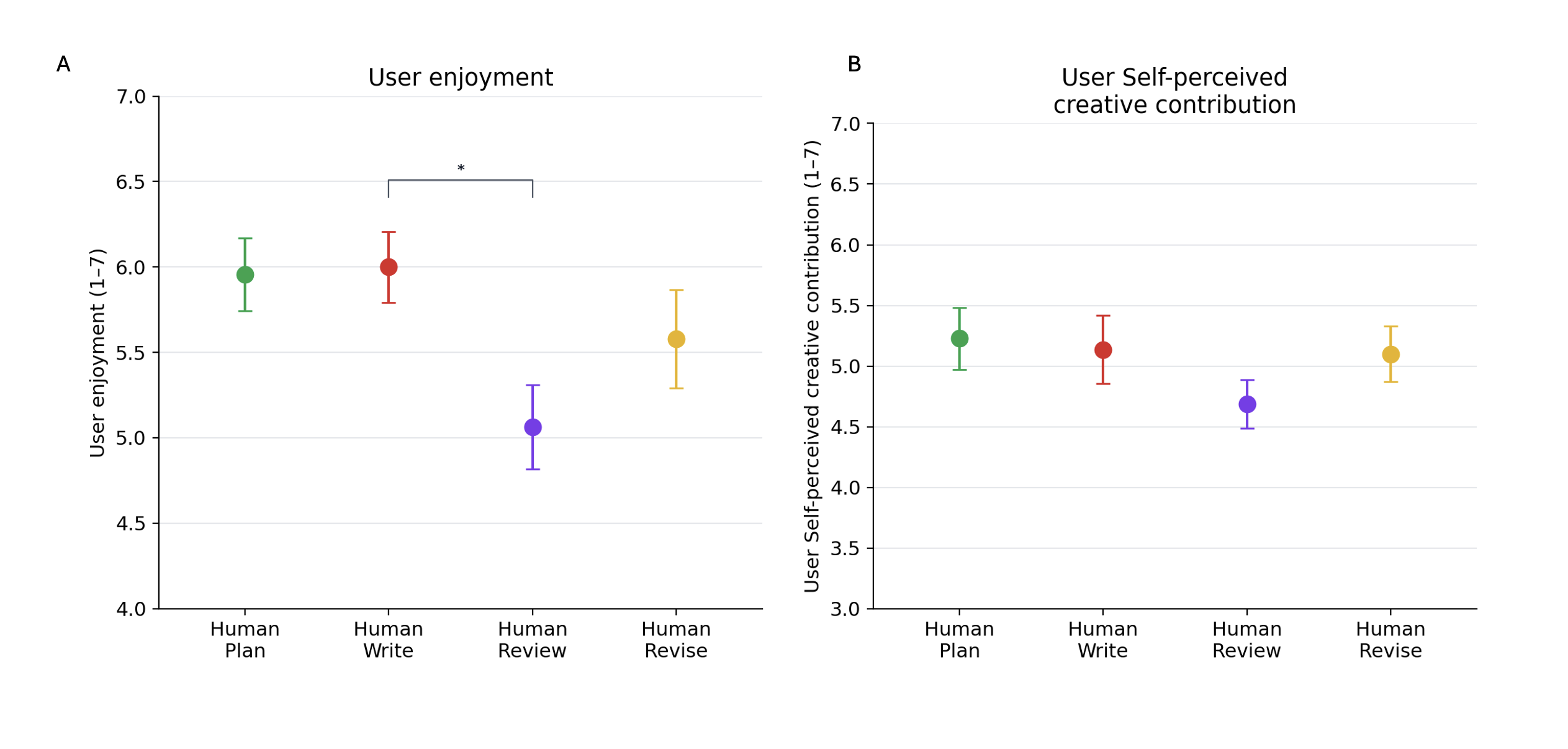}
    \caption{Participant experience by role and condition. A. Self-reported enjoyment. B. Self-perceived creative contribution. Both on 1–7 scales. The shaded region shows each role within Pure Human chains; the unshaded region shows the human role in each Study 2 condition.}
    \label{fig:experience}
\end{figure}

\paragraph{Diversity Dynamics}
Different from creativity dynamics, diversity score changes reliably in every condition, and the direction and magnitude of change differed across conditions.

Human Plan and Human Write follow similar patterns. Both began with higher diversity at Draft than Pure AI (Human Plan: $HL=0.205$, $CI_{95\%}=[0.18, 0.25]$, $\delta=1$, $p<.001$; Human-Write: $HL=0.179$, $CI_{95\%}=[0.16,0.20]$, $\delta=1$, $p<.001$), then declined significantly across rounds (Human-Plan: $\Delta= -0.04$, $CI_{95\%}=[-0.07, -0.02]$, $p<.001$; Human-Write: $\Delta=-0.076$, $CI_{95\%}=[-0.10, -0.06]$, $p<.001$).
This suggests that both a human plan and human writing seed diversity into the story, and the downstream AI stages then homogenize it. Interestingly, Human Plan and Human Write did not differ in diversity at Draft($HL=0.03$, $CI_{95\%}=[-0.01,0.08]$,$\delta=0.30, p=.35$), but the two diverged as AI rounds accumulated. Human Plan was already significantly higher at R1 ($HL=0.05$, $CI_{95\%}=[0.02, 0.09]$, $\delta=0.49$, $p=.02$), and the gap widened by R2($HL=0.08$, $CI_{95\%}=[0.05,0.11]$, $\delta=0.0.739$, $p<.001$). This suggests diversity seeded through a human-written plan is more resilient to AI homogenization than diversity injected by humans writing from an AI plan.
 
Human Review and Human Revise both started with low diversity of AI drafts as expected, then got diversified significantly thereafter. However, Human Review is indistinguishable from Pure AI at every round (all $p=1.0$), indicating that humans only giving feedback adds little diversity to the AI revisions. In contrast, Human Revise gained diversity from Draft to R2 significantly ($\Delta=0.23$, $CI_{95\%}=[0.20,0.25]$, $p<.001$), and this gain exceeded both Pure AI ($HL=0.21$, $CI_{95\%}=[0.19,0.223]$, $\delta=1$, $p<.001$) and Human-Review ($HL= 0.206$, $CI_{95\%}=[0.18, 0.23]$, $\delta=1$, $p<.001$)
This means human revision injects semantic diversity into AI Draft at each round; because no AI stage edits the text afterwards, that injection survives. 

Taken together, diversity enters the chain only through generative human inputs (Human Plan, Human Write, Human Revise) but not evaluation (Human Review). Among generative inputs, diversity survives to the extent the human acts last with no AI stage to follow (Human Revise), and erodes wherever AI processing continues afterward (Human Plan, Human Write).

Across four role placements, Human Plan is the most robust position that maintains high creativity of AI texts and robust diversity of human ideas. Human Write adds the least value, since downstream AI processing washes away its diversity advantage even as it repairs the quality drop. Human Review contributes marginal gains in originality but barely moves diversity. Human Revise adds the most diversity at the cost of creativity.

\paragraph{User experience} 
In the Pure Human and four conditions, we asked participants about their subjective experience after they completed the writing task, including enjoyment of writing process and self-perceived creative contribution. We then computed the scores of each metrics. The enjoyment in Human Review scores significantly lower than Human Write as shown Figure~\ref{fig:experience}. People perceived more enjoyment writing than reviewing ($HL=[0.5, 1.5], CI_{95\%}=[[0.5, 1.5], \delta=0.52, p=.014$). For self-perceived creative contribution, there was no difference among all conditions. These results suggest that humans generally retain their enjoyment and sense of agency when participating in writing chains with AI agents, although participants in the reviewer role reported lower enjoyment.

\begin{figure}
    \centering
    \includegraphics[width=0.9\linewidth]{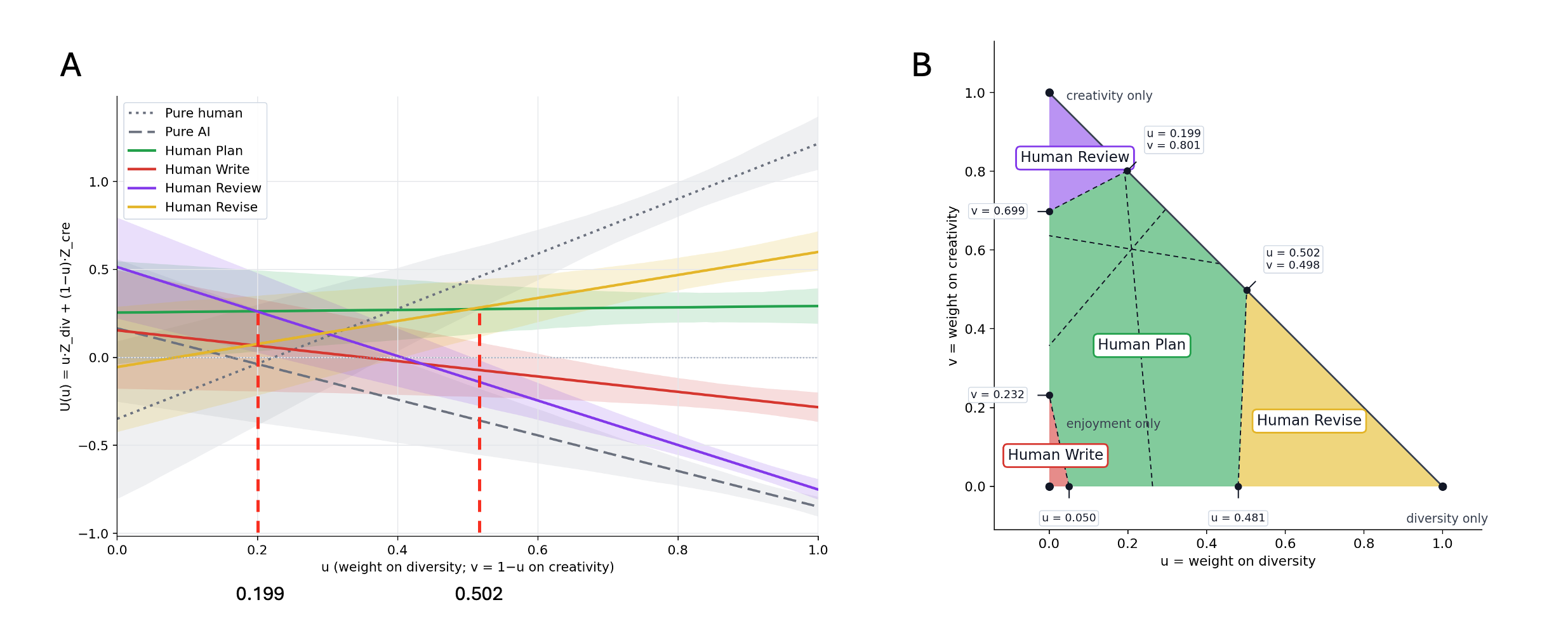}
    \caption{Utility of each human placement at Revision 2 under different weightings of the outcomes. A. Utility $U(u) = u·Z_div + (1- u)Z_{cre}$ as a function of the weight on diversity, $u$. Shaded bands show $95\%$ bootstrap intervals, and vertical dashed lines mark where the utility-maximizing placement changes ($u = 0.199, 0.502$). Pure Human and Pure AI are shown for reference but excluded from the maximization. B. Utility-maximizing placement across all weightings of diversity ($u$), creativity ($v$), and enjoyment ($1 - u -v$). Each point in the triangle is one weighting, colored by the placement with the highest utility.}
    \label{fig:tradeoff}
\end{figure}

\paragraph{Tradeoff between creativity, diversity and enjoyment}
When we jointly consider both creativity, and diversity, the optimal division of labor depends on how much weight is assigned to each outcome. We therefore z-scored the three measures (togehter with the Pure human and Pure AI control) and evaluated all possible weighted combinations of them (Figure~\ref{fig:tradeoff}A). The x-axis represne the tradeoff parmter $u$. When $u=0$ we only consider  creativity and when $u=1$ we only coside diversity. 

%

Human Review maximized utility when diversity received little weight ($u < 0.199$) on diversity and more emphasize on creativity; Human Plan maximized utility at intermediate weights ($0.199 < u < 0.502$), reaching a balance between creativity and diversity; and Human Revise maximized utility when diversity received the majority weight ($u > 0.502$) and put less emphasis on creativity. Human Write never reached the upper utility envelope, meaning Human Write does not serve as an optimal option in any case. This is consistent with our previous results in Section~\ref{sec:study2_result}.
 
We then included enjoyment in the utility function as shown in Figure~\ref{fig:tradeoff}B. 
Each point in the triangle represents a different utility function: locations closer to a vertex place greater weight on that objective, and the color indicates which human role—Plan, Write, Review, or Revise—produces the highest combined utility under that weighting. 
Human planning and human revision dominate most of this space, with the preferred role shifting primarily according to the relative importance of creativity and diversity. Participant enjoyment changes the result only modestly. Although participants particularly enjoyed the writing role, this advantage was offset by lower creativity and diversity, so writing is optimal only in a small region where enjoyment is weighted very heavily. Conversely, review was relatively less enjoyable and rarely optimal.

Take together, Figure~\ref{fig:tradeoff} shows that there is no universally optimal human placement; the preferred role depends on how a system values creativity, diversity, and human experience. At the same time, the utility analysis reinforces the results in Section~\ref{sec:study2_result}: Human Plan emerges as a particularly robust solution to the division-of-labor problem, performing well across a broad range of weightings of all three objectives.


\section{Discussion}



\subsection{Summary of results}
We used transmission-chain experiments with prespecified writing roles—Plan, Write, Review, and Revise—to evaluate creative work in human, AI, and hybrid teams. Creativity was assessed using subjective ratings from non-expert judges, who evaluated the creativity of each story, and diversity using pairwise semantic distances between stories. We also collected participants’ self-reported enjoyment and perceived contribution.

In Study~1, we replicated prior findings that Pure Human chains produced less highly rated but more diverse stories, whereas Pure AI chains produced more highly rated but more homogenized stories. Using LLM-based semantic analysis, we further found that humans introduced new ideas more often but also removed more existing information, whereas AI agents relied more heavily on paraphrasing and elaboration. These differences suggest that the greater diversity of human-generated stories arises in part from distinct writing and rewriting strategies.

In Study~2, we asked a division of labour question: if a human occupies one stage of an otherwise AI-driven pipeline, which role should it be, and how does the answer depend on whether one values creativity, diversity, or the human experience of participation?

All four hybrid conditions converged on similar levels of final creativity, close to those of Pure AI chains. Even when human-written intermediate outputs were initially rated lower, subsequent AI writing or revision could recover creativity. Diversity showed a very different pattern: the four conditions ended at distinct levels spanning much of the range between the Pure Human and Pure AI endpoints. These differences depended on both the human role and the amount of downstream AI processing. Human planning and final revision increased diversity the most; diversity introduced during human writing was largely erased by subsequent AI revision, whereas human review had little effect on diversity. Enjoyment and perceived contribution were largely unaffected by role placement, with the reviewer role as the main exception.

These results suggest that planning may be a particularly effective point for integrating humans into collaborative human–AI systems. Human planning achieved a strong balance across creativity, diversity, and enjoyment, while allowing downstream AI agents to elaborate and refine the initial ideas without eroding its diversity. A second possible strategy is to place humans at the final revision stage, where their contribution cannot be overwritten by subsequent AI processing. This placement increased diversity, but came with some reduction in creativity and lower enjoyment, highlighting a tradeoff between preserving human influence and maintaining output quality and user experience.

\subsection{Homogenization as a Dynamic Force in Human-AI Co-Writing}


Human and AI impose opposite forces to the texts in a sequential writing chain. AI processing pushes drafts toward a high level of creativity across overall dimensions, thereby decreasing diversity in expression and ideas. This normalization of creativity is stable across rounds if no extra forces intervene. AI revisions mostly paraphrase and revise, so they preserve narrative elements while largely homogenizing expression at the semantic level.

In contrast, the placement of human roles enforces diversification in both expression and ideas. Human writers introduce original ideas through diverse and substantial creative operations, yet at the cost of lowering the overall quality of texts toward the Pure Human level. Across rounds, humans transform and integrate elements, resulting in marginal gains in creativity. When humans are placed in different roles in the chain, they are granted different degrees of control over the text, together with different ways of injecting ideas and expressions. The nature of each role also interacts with how humans and AI exert their influence throughout the chain.

The Plan role serves as an intention or guideline for downstream processing. This role has a strong fixation effect on downstream writing. We observe that most elements in the human-written plan are preserved during downstream processing, but diversity still degrades. This may be because AI revision preserves most of the narrative elements proposed by the plan while changing the expression and wording, which leads to lower diversity.

When humans only play the Write role, they are anchored by the AI-written plan. Human operations make major changes to the plot, resulting in a decrease in creativity. However, due to the anchoring effect of the AI plan, their writing has lower diversity than Human Plan. The subsequent AI processing may then normalize the writing toward a high level of creativity again by filling in missing narrative elements and homogenizing expressions through paraphrasing and elaboration.

At later positions of Review and Revise, participants worked on improving a highly rated AI draft. Alternating human and AI in the Review–Revise cycle therefore creates two forces operating at asymmetric positions.

Review grants participants no right to change the text itself, and the extent to which the downstream text changes depends on whether the reviser takes up the feedback. Most of the time, revisers barely take the advice, making it difficult for reviewers to enforce their feedback. On the other hand, revisers hold the final gate for changing the text and are able to inject their own creativity.

When humans serve as reviewers, the strong homogenization of expression in downstream AI processing prevents their feedback from substantially changing the diversity of the stories. Nevertheless, human reviewers are still able to inject originality in terms of ideas.

When humans serve as revisers, there might be two possible consequences. One possibility is that participants are strongly anchored by the previous artifacts because the inferential distance is low. The other is that human participants depart from or even drastically change the previous artifacts. Our results show the latter. They largely drop and transform narrative elements in AI drafts, including both ideas and expressions, which immediately pulls creativity down toward the Pure Human writing level. This is consistent with our observation in Human Write: human forces consistently draw texts toward the creativity level of human writing. This may suggest that the anchoring effect does not override human agency, consistent with our observation that participants perceive a comparable level of creative contribution from humans in the artifacts. Furthermore, by incorporating AI-generated reviews, human revisers can pull creativity back to a comparable level after another round of revisions.

Therefore, the dynamics of creativity and diversity across different conditions are products of the evolving interaction between human and AI forces, as well as the nature of writing roles within the sequential chain.


\subsection{Design implication}


Our results suggest that placing humans at the Plan stage should be a priority when balancing these three dimensions. A human-written plan serves as an executable guideline for the ideas specified in the Writing Prompt, guiding AI writing and downstream processing toward the intended ideas while maintaining high quality and heterogeneity. However, this configuration may risk preserving overly mechanistic expressions or other AI footprints in the resulting artifacts, as suggested by prior literature \cite{chakrabarty2024art}. Beyond creative writing, this configuration may also apply to open-ended workflows that reward differentiation and originality in the market, such as game design. For example, a game designer could specify a design plan and inject creative ideas early in the process, then work with AI coding agents to realize and iteratively refine the design~\cite{DreamGarden}. This configuration also shares a similar core idea with emerging AI-native game designs~\cite{1001Nights,AINativeGamesSurvey}, in which AI agents generate game artifacts on the fly based on human-written scripts and real-time input~\cite{AIDungeon,Dramamancer,Orchid,RuntimeBehaviorGen}. 

A workflow that prioritizes creativity could introduce minimal human intervention during the artifact-generation process, making Human Review a preferable configuration. AI-generated artifacts tend to have heavy AI footprints\cite{kobak2025delving}, which may lead to lower human enjoyment and diversity. However, this trade-off may be less problematic in productive workflows such as coding, where diversity is less important and quality and consistency are often prioritized. This pattern is also common in users' everyday interactions with AI, such as vibe coding \cite{sarkar2025vibe}, where humans initiate a request without specifying a detailed plan, and AI generates the plan and code to fulfill the request. Humans then act primarily as reviewers of the resulting code and output. Future work can test this hypothesis within control codiding experiments \cite{hu2026human}.


Our results suggest that workflows prioritizing diversity may benefit from placing humans in the Reviser role. Positioning a human reviser at the final stage, after AI-generated feedback and before the artifact is completed, gives humans the final opportunity to reshape the output without subsequent AI processing overriding their contribution. This principle may generalize to other creative domains in which humans provide the ``final touch’’ to artifacts otherwise produced through agentic AI workflows, such as visual art or design.

When human enjoyment is the primary consideration, humans should serve primarily as executors, here represented by the Writer role. This may be because humans enjoy realizing their ideas through creative actions while being scaffolded by AI-generated guidelines; however, whether, where, and why people find certain parts of the writing process enjoyable are grounds for future work especially as we consider what might make a collaboration meaningful~\citep{collins2026meaningful}. The AI Planner provides a framework that reduces the cognitive burden of constructing an entire story from scratch, while still leaving substantial freedom for execution. Therefore, human writers can introduce their own ideas without being constrained by specific expressions.

As LLM-generated content becomes cheaper and more abundant, human attention increasingly becomes the limiting resource in human–AI co-creative pipelines. Deciding where to spend that scarce attention is therefore a design question that extends well beyond the four roles we study here, and one that merits systematic investigation across a wide range of creative and productive domains. Our work focused specifically on creative writing, but similar methods could be applied to other open-ended creative domains, such as music and drawing \cite{tchernichovski2025constraints}, as well as domains with objective performance criteria, such as semantic search \cite{li2026human} and coding \cite{hu2026human}.

\subsection{Limitations and Future Work}


Creativity is inherently difficult to measure. Our study operationalizes it as a quantitative rating collected in a controlled lab setting, with non-expert writers of varying writing experience producing the stories. This approach carries several limitations.
First, we relied on non-expert raters to judge creativity. Prior work shows that rating quality is closely tied to rater expertise, suggesting expert rating is more reliable than non-expert \cite{chakrabarty2024art,karpinska2021perils}. This may limit the validity of our creativity measure. That said, this setup mirrors everyday practice: most creative writing is produced for, and judged by, non-expert audiences rather than literary professionals.
Second, our writers were themselves non-experts with varying levels of writing experience, and effort likely varied across participants. Separately, prior research shows that expert-written stories are still rated as more creative than AI-generated ones\cite{chakrabarty2024art}. Our findings may therefore not generalize to professional creative industries; we acknowledge this and scope our claims to common, everyday creative practice rather than professional writing. Future work could test how the creativity–diversity tradeoff changes when both the writers and the raters assessing creativity are experts. 

Our study takes the four Flower and Hayes (1981) writing processes \cite{flower1981cognitive} — planning, translating, and reviewing — as the basis for our pipeline, but operationalizes them as a strict, non-overlapping sequence: Plan, then Write, then Review, then Revise. Our design assumes a flatter process in which each stage happens once, in order. But these processes are recursive and interleaved \cite{flower1981cognitive}. There is empirical support for something like the pantser mode in a ``dual-process'' account of writing \cite{baaijen2018discovery}, in which top-down expansion of a high-level plan is closely intertwined with bottom-up, sentence-level text generation. Our pipeline does not capture this bottom-up mode, which, anecdotally, may be the more common way people actually write.How to support bottom-up, moment-to-moment writing in human-AI collaboration is, we think, a valuable direction for future work.

We then discussed the limitations of AI models we used, in terms of model parameters, prompt engineering and model choices. Our study used a single AI model (GPT‑4 Turbo) with a fixed temperature (T = 0.5) and a constant seed across generations, which may have made model behavior more deterministic and suppressed output diversity. To probe this, we reran the pipeline with T = 1 and a new random seed for each API call (Supplementary Figure 2). Diversity scores under this higher-randomness setting exceeded Pure AI at every round but remained below Pure Human at every round, consistent with our Study 1 finding. This suggests our Study 2 conclusions are largely insensitive to this choice, since they rest on the ordinal relationship between conditions (AI diversity < Human diversity) rather than on the exact magnitude of AI diversity, and this ordering holds under higher temperature. Higher randomness could plausibly raise creativity by promoting more divergent operations in the output. However, temperature has been shown to correlate only weakly with novelty and moderately with incoherence \cite{peeperkorn2024temperature}. This suggests higher temperature may buy diversity at the cost of coherence, which would in turn shape how the human role interacts with the creativity dynamics we report. We leave the investigation for this tradeoff to future work.

A second common way to introduce randomness is multi-persona prompting, where each AI instance is assigned a distinct persona sampled from PersonaHub database\cite{ge2024scaling}. We tried this setup too, and it also produced diversity scores higher than Pure AI but lower than Pure Human (Supplementary Figure 2). However, we cannot exclude the possibility that alternative prompt-engineering techniques could further increase diversity. We leave the systematic exploration of such approaches to future work.

We used only one model family. In particular, we did not test open-weight models, which would afford greater reproducibility, or models from different providers, which may exhibit different creativity–diversity tradeoffs. However, recent work suggests LLMs across companies and parameter scales tend to produce more homogeneous outputs than humans and behave similarly to one another \cite{wenger2025we,jiang2026artificial}. ecause the PsyNet framework can seamlessly integrate different models, future work can systematically test whether these findings generalize across a broader range of model families and providers.


%

\section{Conclusion}
Our work examined where human contribution should enter in an AI-led creative workflow. We approached this question with a staged transmission-chain paradigm that decomposes story writing into Plan, Write, Review, and Revise, and assigns a single role to humans within an otherwise AI chain. We found that human contributions interact with AI homogenization effect in different role placement, resulting in tradeoff between creativity and diversity. We hope this work encourages designers of human–AI creative pipelines to treat the placement of human work as an important design variable, and to evaluate it at the level of the collection as well as the individual artifact.

\begin{acks}
To Buchi, my lovely paw friend.
\end{acks}



\bibliographystyle{ACM-Reference-Format}
\bibliography{reference}

\appendix

\end{document}